\documentclass{INTERSPEECH2023}
\usepackage{multirow}
\usepackage{flushend}

\usepackage{graphicx}
\usepackage{ulem}
\usepackage{xcolor}

\definecolor{green}{RGB}{10,150,35}

\interspeechcameraready 

\title{Masking Kernel for Learning Energy-Efficient Representations for Speaker Recognition and Mobile Health}

\name{Apiwat Ditthapron, Emmanuel O. Agu, Adam C. Lammert \thanks{This material is based on research funded by DARPA under agreement number FA8750-18-2-0077.}}
\address{Worcester Polytechnic Institute, 100 Institute Rd, Worcester, MA 01609 ,USA}
\email{aditthapron@wpi.edu, emmanuel@wpi.edu, alammert@wpi.edu}

\begin{document}

\maketitle
 
\begin{abstract}

Modern smartphones possess hardware for audio acquisition and to perform speech processing tasks such as speaker recognition and health assessment. However, energy consumption remains a concern, especially for resource-intensive DNNs. Prior work has improved the DNN energy efficiency by utilizing a compact model or reducing the dimensions of speech features. Both approaches reduced energy consumption during DNN inference but not during speech acquisition. This paper proposes using a masking kernel integrated into gradient descent during DNN training to learn the most energy-efficient speech length and sampling rate for windowing, a common step for sample construction. To determine the most energy-optimal parameters, a masking function with non-zero derivatives was combined with a low-pass filter. The proposed approach minimizes the energy consumption of both data collection and inference by 57\%, and is competitive with speaker recognition and traumatic brain injury detection baselines.

\end{abstract}

\section{Introduction}

Speech processing hardware embedded into smartphones facilitates on-device performance of speech tasks such as voice authentication and health assessment, either as short episodic sessions or continuously. Most state-of-the-art speech processing pipelines utilize Deep Neural Networks (DNNs) to achieve accurate analyses, with inference typically done either locally on the mobile device or on a remote server. On-device DNN inference preserves the speaker's privacy more than remote inference but requires audio to be transmitted to the server, which consumes additional energy.  

Although smartphones are now powerful enough to perform real-time DNN inference, energy consumption remains an issue especially when high-performance DNNs are used for continuous, passive health assessment~\cite{ditthapron2022continuous}. Prior mobile speech processing proposed improving energy efficiency by using compact DNN models~\cite{nunes2020mobilenet1d}, or energy-efficient hardware~\cite{wang2014vlsi,lu2011speakersense} and feature extraction approaches~\cite{bergsma22_interspeech}. DNN model complexity can be reduced by up to 86\% using audio features such as Mel-frequency cepstral coefficients (MFCCs), or by factorizing the DNN model into smaller kernels using depth-wise convolution~\cite{nunes2020mobilenet1d}. Windowing is a common speech processing step, in which the input signal is divided into temporal segments for feature extraction and DNN. Inspired by the feature compression approach, in this paper, we propose reducing DNN input dimensions by identifying the smallest usable window length and sampling rate, which in turn reduces the energy consumed by audio data acquisition and DNN inference. Our proposed method is a signal pre-processing step that can be integrated into most speech processing pipelines including compact speaker recognition and health assessment DNNs.

\begin{figure}[!tb]
    \centering
\includegraphics[clip,trim=0.1cm 5cm 7.5cm 0cm,width=1\linewidth]{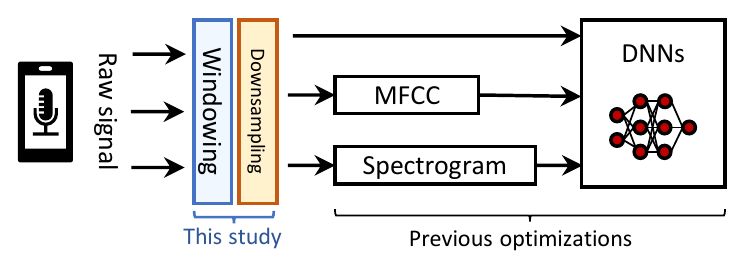}
    \caption{Common speech processing pipeline. This study introduces learnable masking into windowing and downsampling to reduce computational complexity in downstream processing. The method can be used with speech features and DNNs previously proposed for resource-constrained computing.}
    \label{fig:speech_processing_pipeline}

\end{figure}

As illustrated in Figure~\ref{fig:speech_processing_pipeline}, DNNs for speech processing typically operate on a sequence of discrete signal \textit{chunks}, which are generated during pre-processing steps performed before feature extraction. Each chunk contains $n\times s$ frames from $n$ seconds of audio sampled at $s$ Hz and has a length that varies depending on the speech task (e.g., 200 milliseconds (ms) for speech recognition and up to 15 seconds for depression detection~\cite{ravanelli2018speaker,chlasta2019automated}). While recording and processing longer speech chunks sampled at higher sampling frequencies improve recognition performance~\cite{lu2011speakersense}, this approach also consumes higher energy, which limits its practical application~\cite{oletic2018system}. The optimal size of an audio chunk is typically determined as a hyperparameter using grid search or Bayesian optimization during DNN model training~\cite{lu2011speakersense,victoria2021automatic}. 

To optimize the shape of the input signal and derive energy-efficient parameters, we propose determining the most energy-efficient speech duration $m$ and sampling rate $s$ via masking during DNN training while also learning the parameter $\theta$, where $\theta$ are weights in the backbone DNNs. Additional windowing and down-sampling layers for optimizing $m$ and $s$, as well as $\theta$ are proposed to be included at the beginning of the DNN. We envision that such parameter learning will be done during training on a remote server with inference running locally on a mobile device. Gaussian, Hamming, Hann, and Tukey windows~\cite{bloomfield2004fourier} were evaluated as masking functions during back-propagation in order to learn $m$ in the windowing layer. In contrast with the traditional use of masking (or soft-masking), we applied a binary hard mask step for constructing a discrete window. This binary step is used in the down-sampling layer as a masking function for learning the appropriate discrete Fourier transform signal bandwidth.

Our approach is inspired by prior learning approaches that discover an optimal end-to-end DNN architecture, such as  
Neural Architecture Search (NAS) ~\cite{romero2022flexconv,riad2022learning}. Utilizing only training data, NAS transforms each layer of the DNN architecture into derivable functions, such as masking, which can be back-propagated via gradient descent. Flexconv~\cite{romero2022flexconv} proposed learning the optimal kernel size for the image recognition task by using a Gaussian function as a mask on the convolution weights. DiffStride~\cite{riad2022learning} proposed masking for back-propagation in order to learn the scaling factor of the pooling layer. Searching for optimal architectures using NAS achieves performance superior to hyper-parameter tuning. The use of masking in previous work is similar to ours, but the main distinctions are in the learning objective and methodology. Our method applies masking to the input, as a signal pre-processing, and not on the weights to optimize the model architecture as in NAS.

An energy-efficient penalty is introduced to prevent $m$ and $s$ from expanding, reducing the amount of energy required for inference and data recording on a linear scale~\cite{oletic2018system}. Our proposed method is able to reduce the energy utilized for inference while minimizing losses in performance. We evaluated the windowing layer, down-sampling layer, and energy-efficient penalty at the window level (one speech chunk) and sentence level (multiple speech chunks) for the speaker recognition task, and for the continuous Traumatic Brain Injury (TBI) detection task, which are the common tasks in mobile health. The energy used for DNN inference significantly improved in all three scenarios, whereas energy expenditure during data acquisition was reduced only in the first scenario. We also show that the parameters learned in the windowing layer are compatible with the compact DNN model and compressed features and outperform the parameters obtained from traditional hyperparameter tuning, improving both accuracy and power efficiency.

\section{Proposed method}
\label{sec:proposed_method}

The optimization of window size and the sampling rate is accomplished via back-propagation through windowing $\mathcal{W}_{m}$ and down-sampling layers $\mathcal{D}_{s}$. The parameters in these two layers ($m,s$) are learned jointly with the ($\theta$) parameter in the DNN but are controlled by an energy-efficiency penalty $\mathcal{J}$ in order to minimize the size of the speech sample. Given a speech model $\mathcal{F}_{\theta}(x)$ with a loss function of $\mathcal{L}(x,y)$, parameters are optimized from dataset $\{x_i,y_i\}_{i=0}^P$ by 
$\mathrm{argmin}_{\theta,m,s} \sum_{i=0}^P \mathcal{L}\big(\mathcal{F}_{\theta}( \mathcal{W}_{m}(\mathcal{D}_s(x_i)),y_i\big) + \mathcal{J}(m,s)$

\textbf{Windowing layer}:
Let $x_i\in \mathbb{R}^N$ be a speech sample, composed of $N$ frames where $N$ is the upper bound of window length. Windowing allows a signal of length $m$ ($\lfloor\frac{N-m}{2}\rfloor \leq n\leq \lfloor\frac{N+m}{2}\rfloor$) into the DNN. To learn $m$ via gradient descent, derivatives of the masking function must be non-zeros. A rectangular window, a standard method for segmenting the signal for DNNs, is defined to have values of 1 within length $m$ and values of 0 everywhere else, resulting in zero derivatives.  This study proposes hard-masking, a learnable rectangle window, that uses functions with a peak at its center during back-propagation. Masking functions considered are the well-known Gaussian, Hamming, Hann and Tukey functions~\cite{bloomfield2004fourier}. 

The Gaussian window function has a mean value of $\lfloor \frac{N-1}{2}\rfloor$ with a learnable variance $\sigma^2$. This study defines $\sigma^2$ in terms of window $m$ at which the function is approaching zero ($m^2=-8\log(\epsilon)\sigma^2, \epsilon =1e^{-5}$) as  
$w_G(n;m) = exp(4\log(\epsilon)(n-\lfloor \frac{N-1}{2} \rfloor)^2/m^2)$.
Hamming and Hann windows are defined as $w_{HM}(n;m) = 0.54-0.46\cos(2\pi (n-\lfloor \frac{N-m}{2} \rfloor)/(m-1))$ and $w_{HN}(n;m) = 0.5-0.5 \cos(2\pi (n-\lfloor \frac{N-m}{2} \rfloor)/(m-1))$, respectively. Tukey is also included as a tapered cosine function of $w_{HN}$.

A window $w(n;m)$ is applied to $x(n)$ to attenuate values outside the window. We call the output of this operation soft-masking and consider it a baseline evaluation method. To create hard-masking $\mathcal{W}_m$, a value of 1 is assigned to non-zero values of soft-masking. The hard-masking derivative of $\delta \mathcal{L}/\delta m$ is computed by applying a straight-through estimator~\cite{bengio2013estimating} to $w$.

\begin{figure}[!tb]
    \centering
\includegraphics[clip,trim=0cm 0cm 0cm 0cm,width=1\linewidth]{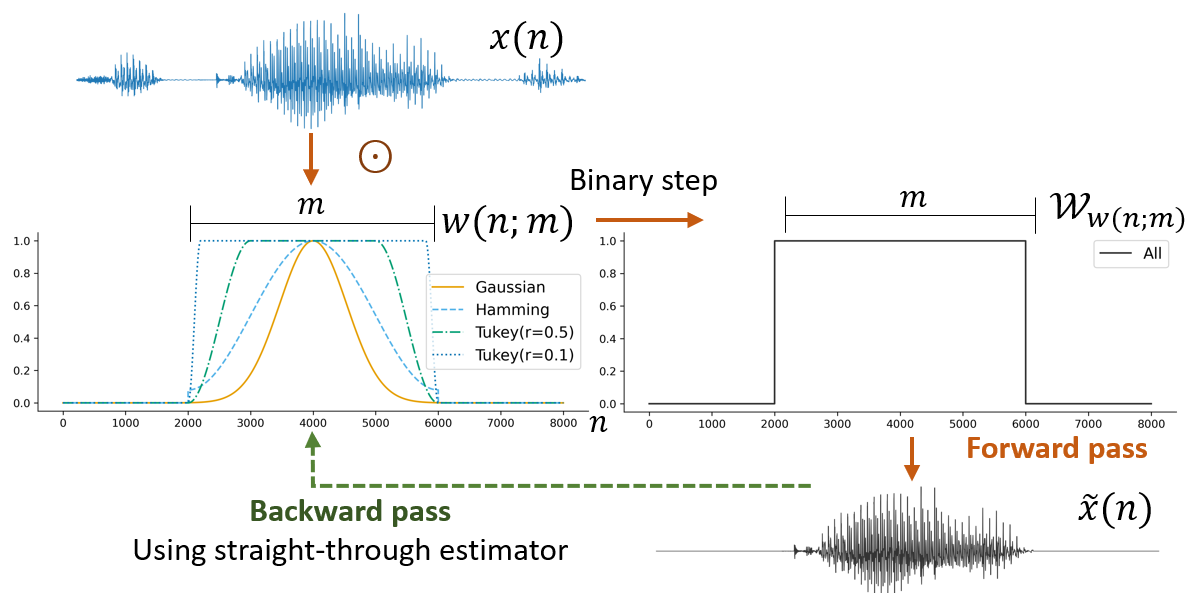}
    \caption{Windowing layer using hard-masking}
    \label{fig:windowing}
\end{figure}

\begin{figure}[!tb]
    \centering
    \includegraphics[clip,trim=0cm 0cm 0cm 0cm,width=1\linewidth]{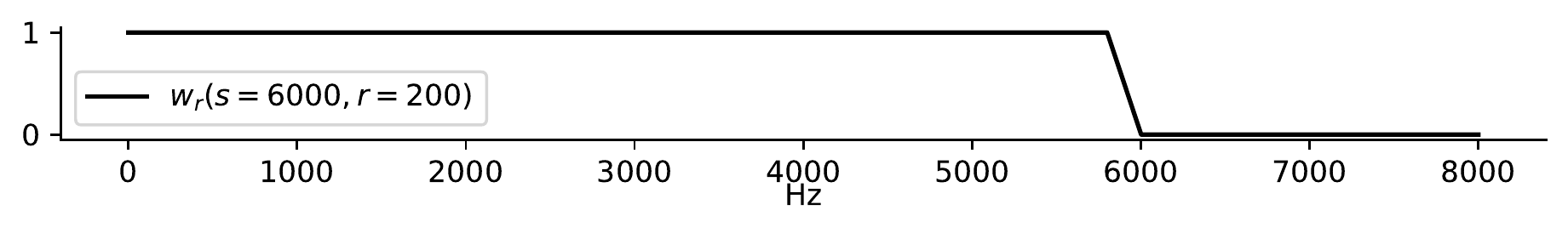}
    \caption{Masking $w_r$ in down-sampling layer}
    \label{fig:down_sampling}
\end{figure}

\textbf{Down-sampling layer:} The down-sampling layer $\mathcal{D}_s$ applies masking in the frequency domain, resampling $x$ to $2s$ Hz. The discrete Fourier transform $X(\hat{n}) = FFT(x(n))$ is obtained from the Fast Fourier Transform (FFT) of $x(n)$. Due to Hermitian symmetry, the term with negative frequency can be disregarded. A rectangle mask is used as a low-pass filter to zero frequency bins higher than $s$. To reduce artifacts from the rectangle mask and allow back-propagation, a linear function is applied, which extends the cutoff frequency by $r$. The mask $w_r(n;s,r)$ is defined as $min(1,max(-\frac{n-s}{r})), 0 \leq n\leq N $
,visualized in Figure ~\ref{fig:down_sampling}. After applying $w_r(n;s,r)$ to $X(\hat{n})$, $x(n)$ is downsampled to $s$ Hz using inverse FFT only on DFT bins between $0$ and $s$ Hz, mathematically explained by $ \mathcal{D}_s= iFFT(X(\hat{n}) \odot w_r(\hat{n};s,r))$, where $0<\hat{n}\leq s$.

\textbf{Energy-efficient penalty:}
A penalty term is introduced to minimize window length and sampling rate, which, in turn, reduces the energy required for data acquisition and inference. The energy-efficient penalty $\mathcal{J}(m,s) = \lambda \big [ \frac{\max( m-\mu_m,0)}{\mu_m} +\frac{\max( s-\mu_s,0)}{\mu_s} \big] \bar{\mathcal{L}}$ is incorporated into the loss function to penalize $\mathcal{L}$ if $m$ or $s$ increases from their average values ($\mu_m$,$\mu_s$) in the preceding epoch. The penalty values are normalized and added proportionally to the value of $\bar{\mathcal{L}}$ (no gradient).
$\lambda$ is adjustable to control the penalty term.  $\mathcal{J}$ is clipped at zero to prevent an exploding gradient.

\section{Evaluation}
\label{sec:experiment}

The proposed method was evaluated using state-of-the-art DNNs previously proposed for speaker recognition (short) and TBI detection (continuous, passive health assessment) tasks. Our implementation is publicly available at \url{https://github.com/aditthapron/windowMasking}.

\subsection{Speaker recognition task}
The speaker recognition speech processing task tries to identify a speaker based on their voice characteristics. On smartphones, speaker recognition is frequently performed as continuous authentication, consuming significant energy~\cite{bouraoui2017hardware}.

\textbf{Dataset}: Text-independent speech from the TIMIT corpus was used to train and evaluate the model~\cite{garofolo1993darpa}. Read speech in English was collected from 462 speakers in 16-bits with a sampling rate of 16 kHz.
All data pre-processing steps, including removing non-speech segments, removing calibration sentences, and normalizing the amplitude, were performed similarly to ~\cite{ravanelli2018speaker}. The space between each window center was fixed at 10ms. $M$ was set to 500ms. The split between training and testing was the same as in ~\cite{ravanelli2018speaker}.

\textbf{Evaluation Metric:} Classification Error Rate (CER) is reported at both the window and sentence levels. At the window level, the speaker with the highest negative log-likelihood is predicted, whereas, the negative log-likelihood from all windows is summed to make the prediction at the sentence level. Reduction of $m$ in window-level speaker recognition implies a reduction in the duration of speech necessary to collect. To assess training consistency, all evaluations were repeated ten times with random seeds of varying values.

\textbf{DNN architecture and features:} The proposed method was evaluated for speaker recognition tasks using two DNNs, SincNet~\cite{ravanelli2018speaker} and Am-MobileNet~\cite{nunes2020mobilenet1d}. Raw audio was input to both models, whereas MFCC features were input to Am-MobileNet.
SincNet~\cite{ravanelli2018speaker} replaced traditional convolutional weights with the Sinc function as the kernel in CNN layers. The model consists of one CNN layer with Sinc filters and two conventional CNN layers. After the CNN, the tensor is transformed into a one-dimensional tensor to classify the speaker. SincNet can only apply to raw audio because of the Sinc layer.
Am-MobileNet~\cite{nunes2020mobilenet1d} adapted the MobileNetV2 model \cite{sandler2018mobilenetv2}, which uses depthwise convolution and an inverted Residual Block to improve model efficiency, for the speaker recognition task. The Additive Margin (AM) was also introduced into the Softmax activation function to improve the separation margin between the decision boundary of the speaker class. For MFCC features, the first 40 Mel bands were extracted using Librosa~\cite{mcfee2015librosa}. Due to the short duration of the signal, the length of the FFT window was reduced to 1024 and the hop length to 128. Thirteen MFCCs were then extracted from Mel-spectrograms. Results with delta MFCCs were not reported as there was no performance gain.

\textbf{Experiment:} We extended SincNet~\cite{ravanelli2018speaker} to learn energy-efficient parameters by including windowing and $\mathcal{D}_s$ layers prior to the SincNet layers. As the input shape changes throughout the learning process, the layers following the CNN were modified to only apply weights to the signal's valid length.

\subsection{TBI detection task}
Frequently, impaired speech is considered a TBI biomarker that can be detected via continous speech assessment using a DNN running on a smartphone, preventing fatalities and facilitating the recovery of TBI~\cite{ditthapron2022continuous}. 

\textbf{Dataset:} Speech from the Coelho TBI corpus~\cite{coelho2002conversational}, was used for evaluation.
The Coelho corpus contains speech during story retelling, story generation, and conversation discourses from 55 subjects with non-penetrating head injuries and 52 subjects without head injuries. We used the speech collected during the conversation discourses for evaluation. Pre-processing steps from~\cite{ditthapron2022continuous}, included 1) removing noisy audio, 2) normalizing audio magnitude, and 3) vocal-tract length normalization. The speech was recorded at a sampling rate of 44.1 kHz, where \cite{ditthapron2022continuous} down-sampled the signal to 16 kHz. This study initialized $s$ in the down-sampling layer to 22 kHz.

\textbf{Metric:} Balanced Accuracy ,$(\text{Sensitivity}+\text{Specificity})/2$, is reported using subject-level split 10-fold cross-validation.

\textbf{DNN architecture:} The cascading Gated Recurrent Unit (cGRU) previously proposed for TBI detection from speech~\cite{ditthapron2022continuous} was the DNN model. cGRU is a two-step DNN where the first model extracts TBI features from 200 ms of speech using five CNN layers, and the second model applies a GRU on stacked features from the first model for binary TBI classification. The CNNs were applied on 200 ms with an interval of 25 ms, and the GRU makes a TBI prediction over 4 s of speech.

\textbf{Experiment:} We integrated the proposed method into the cascading Gated Recurrent Unit (cGRU) model~\cite{ditthapron2022continuous} to learn an energy-efficient input for the TBI detection task. Windowing and down-sampling layers were applied at the instance level ($M =$ 8 s). Due to space constraints, ablation results are only reported for speaker recognition.

\subsection{Model optimization Baselines and metrics}
Grid search and Bayesian model-based optimization were commonly used to tune hyperparameters, including windowing parameters~\cite{he2021automl}. In this study, optimal $m$ and $s$ were searched in ranges of [100,300] ms and [6000,8000] Hz for speaker recognition, and [2,8] s and [6000,22050] Hz for TBI detection using grid-search. Ten evenly spaced values were used for each parameter interval. In Bayesian model-based optimization, the Tree-structured Parzen Estimator (TPE)~\cite{bergstra2011algorithms} was utilized. TPE uses past evaluations of hyperparameters to construct a probabilistic model over multiple iterations. We also considered a low-pass filter with a learnable Sinc filter~\cite{ravanelli2018speaker} as a baseline for $\mathcal{D}_s$ layer. 

\textbf{Energy-efficient metrics:} The numbers of parameters and Multiply-Accumulates (MACs) have previously been shown to be effective estimations of DNN energy consumption at inference~\cite{sze2017efficient}. Our proposed method adds window length parameters $m$ and sampling rate $s$ to the model, but MACs change significantly depending on the size of the input ($m\times s$). Energy consumption at inference can be reduced by lowering the MAC, whereas power consumption during speech recording can be reduced by lowering $s$. Average inference time, measured on the Samsung Galaxy S22 device, and average training time, measured on the Nvidia Tesla V100, are reported. 

\section{Result}
\label{sec:result}
\subsection{Speaker recognition} 
\begin{figure}[!tb]
    \centering
    \includegraphics[clip,trim=0cm 0.2cm 0cm 0.2cm,width=1\linewidth]{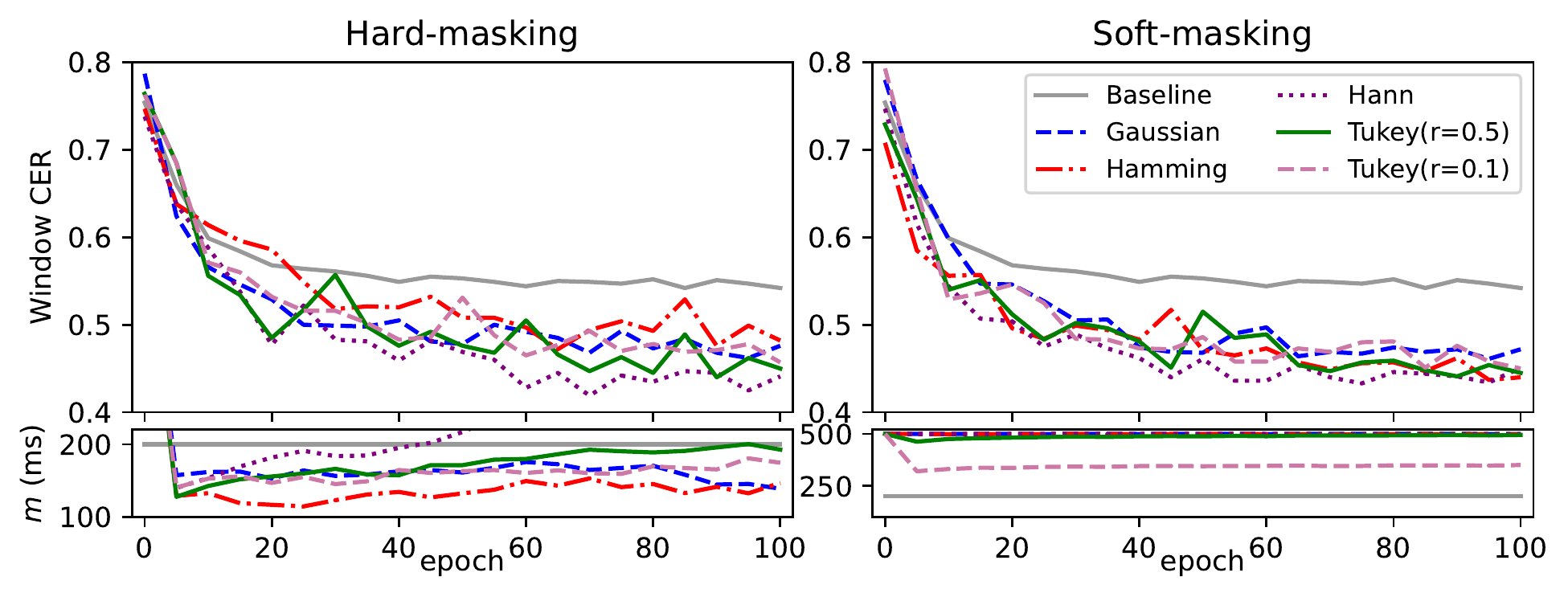}
    \includegraphics[clip,trim=0cm 0.2cm 0cm 0.2cm,width=1\linewidth]{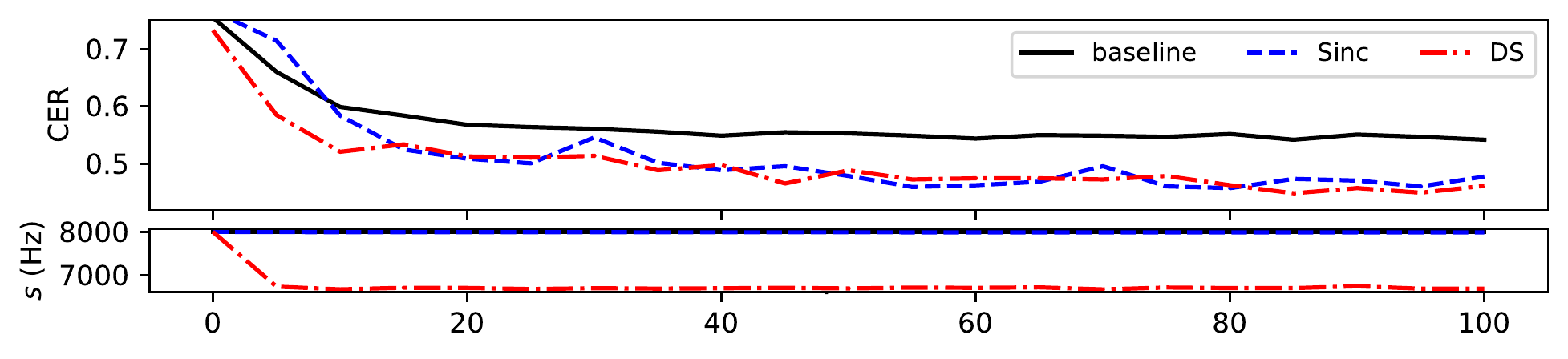}
    \caption{\textbf{TOP}: Window-CER between hard-masking and soft-masing, \textbf{BOTTOM}: Window-CER using Down-sampling layer (DS) and Sinc filter}
    \label{fig:mask}
\end{figure}

\begin{figure}[!tb]
    \centering
    \includegraphics[clip,trim=0cm 0.2cm 0cm 0.2cm,width=1\linewidth]{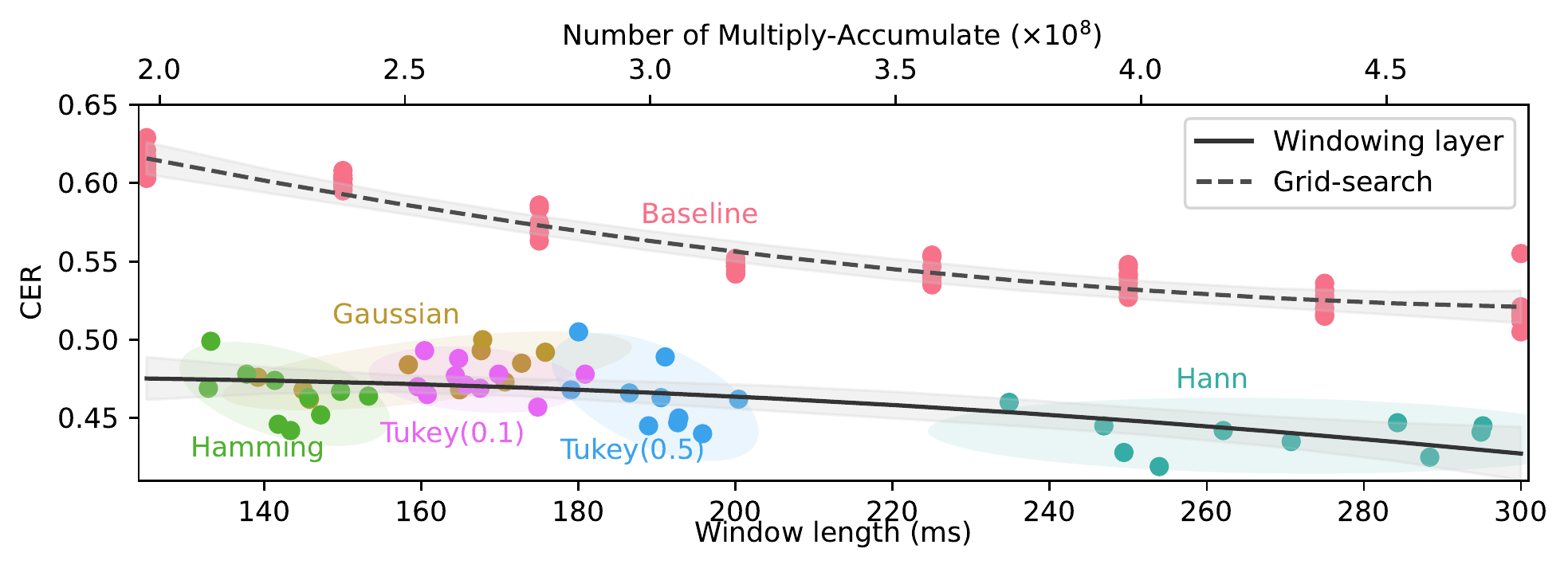}
    \caption{Trade-off between CER and window length}
    \label{fig:trade-off}
\end{figure}

\begin{figure}[!tb]
    \centering
    \includegraphics[clip,trim=0.2cm 0cm 0cm 0.2cm,width=1\linewidth]{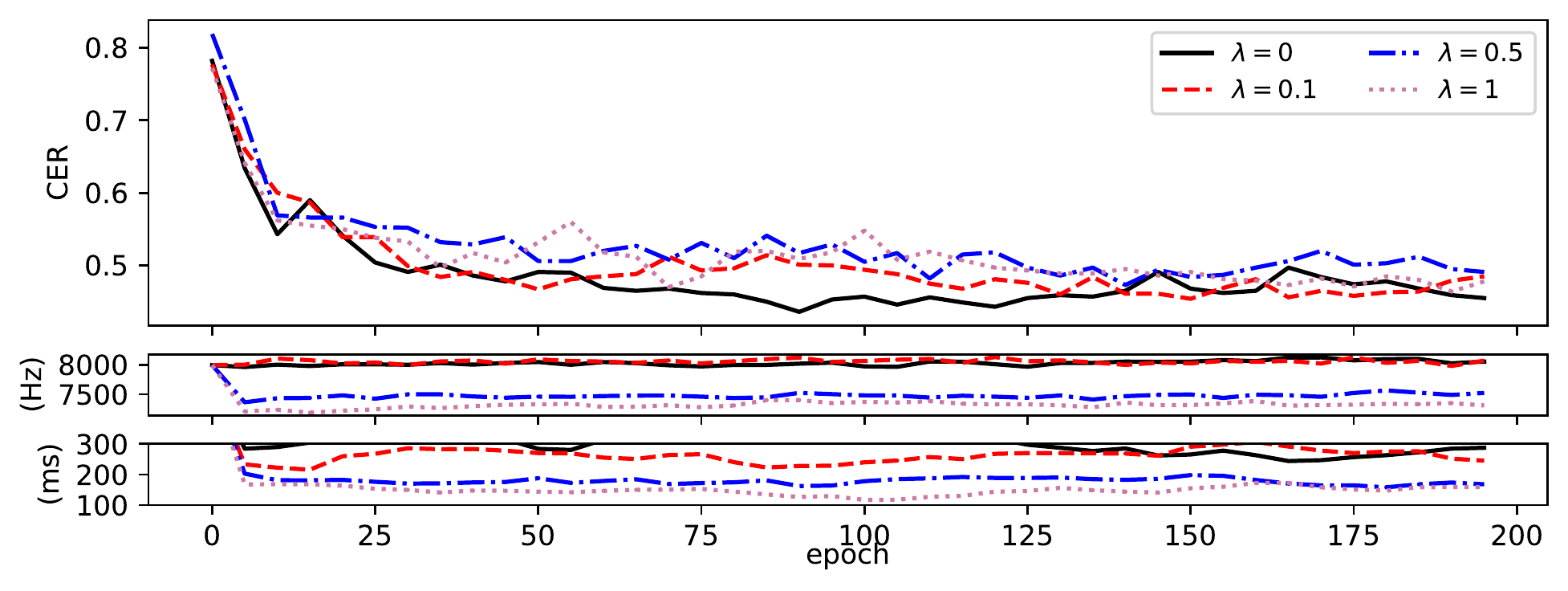}
    \caption{Effect of energy-efficient penalty}
    \label{fig:penalty}
\end{figure}

Windowing functions are compared as hard-masking and soft-masking for speaker recognition in Figure ~\ref{fig:mask} (top). Only the hard-mask is able to optimize $m$, approximately at 120-200 ms. Window-CER is significantly lower than the baseline trained on a 200-ms speech. From the plot between CER and window length (Figure ~\ref{fig:trade-off}), Hamming window was the most efficient at reducing window length while maintaining the same error range as Gaussian and Tukey windows. The CER of the Hann window is lower than the other windows, however, using $m$ higher than 200ms. Figure~\ref{fig:mask} (bottom) compares the $\mathcal{D}_s$ layer to the Sinc. $\mathcal{D}_s$ is competitive with the Sinc filter but with a significantly lower sampling rate of 7.2 kHz. Together, the two proposed masking layers can be optimized using penalty terms, as shown in Figure ~\ref{fig:penalty}. $\lambda=0.5$ and $1$ provide the most energy-efficient performance, reducing the window length to 118 ms and sampling rate to 7.2 kHz while maintaining a CER of 0.49. 

Table~\ref{table:speaker_recognition} shows comparisons between the proposed method, grid-search, and TPE. For raw audio, the proposed method is capable of reducing MAC by 73\%, with comparable performance using Am-MobileNet. The energy used for data acquisition is also reduced by up to 57\% for window-level speaker recognition. In the SincNet model, the proposed methods reduced the MAC in the model by 49\% with a performance gain in window-level CER. Our proposed method is able to improve energy efficiency and CER in both models. However, CER and the energy utilized by SincNet are significantly higher than for AM-MobileNet, which is due to the DNN architecture itself.

To further improve power efficiency, the proposed method was evaluated on MFCC features using AM-MobileNet. The inference time, including MFCC extraction, was reduced from 12.7 ms to 8.2 ms with competitive CERs. Although speaker CERs using MFCC are higher than for raw audio, energy and time used at inference are reduced by half.

In most experimental setups, speaker CERs using our proposed method are lower than the baselines. The improved CERs may be due to the mechanism of optimizing the windowing layer, which allows other parameters in the model to learn on various receptive fields over the training epochs. This conjecture is evidenced by the result of the fixed value, which trains the model using fixed $m$ and $s$ values as in the proposed method, which has inferior results.

\begin{table}[!tp]
    \centering
    \caption{Speaker recognition and TBI detection results }
    \resizebox{\columnwidth}{!}{
    \addtolength{\tabcolsep}{-0.2em}
    \begin{tabular}{lll|cc|ccccc}
        \toprule
         \multicolumn{3}{c|}{\textbf{Speaker}}& \multicolumn{2}{c|}{\textbf{CER} $(\%_{\pm std})$} & \multicolumn{5}{c}{\textbf{Energy-efficient metrics}} \\
         \multicolumn{3}{c|}{\textbf{classification}}& Window& Sentence & $m$ & $s$ & MAC & Inference & Training\\
         &&&&& (ms) & (Hz) & & Time (ms) & Time\\
        \midrule
        \multirow{5}{*}{{\rotatebox[origin=c]{90}{\textbf{SincNet}}}} & \multirow{5}{*}{{\rotatebox[origin=c]{90}{(Raw audio)}}}
        & $\mathcal{W}_{HM}$ & $\textbf{48.6}_{.4}$& $1.02_{.13}$ & \textbf{118} & 8k & 0.58 & 118 & 0.96  \\
        & & $\mathcal{W}_{HM}$ + $\mathcal{D}_s$ & $\textbf{49.1}_{.3}$ & $1.08_{.11}$& 120 & 7.2k  & \textbf{0.51 } & \textbf{110} & 1.05\\
        
        & & Grid-search & $53.8_{.1}$&$\textbf{0.94}_{.09}$& 200& 8k & 1 & 184 & 1 \\
        & & TPE & $52.8_{.7}$ & $1.29_{.13}$ & 272 &7.6k & 0.87 & 135 & 22\\
        & & Fixed values & $56.0_{.1}$& $1.24_{.17}$ & 120 & 7.2k & 0.51 & 118 & 0.64  \\
        
        \midrule
        
        \multirow{10}{*}{{\rotatebox[origin=c]{90}{\textbf{Am-MobileNet}}}} & \multirow{5}{*}{{\rotatebox[origin=c]{90}{(Raw audio)}}}
        & $\mathcal{W}_G$  & $\textbf{21.9}_{.1}$ & $0.36_{.13}$ & 106 & 8k & 0.46 & 16.7 & 1.13 \\
        & & $\mathcal{W}_G$ + $\mathcal{D}_s$ & $22.8_{.2}$ & \textbf{$\textbf{0.32}_{.18}$} &\textbf{99} & \textbf{5.1k} & \textbf{0.27} & \textbf{16.1} & 1.02\\
        
        & & Grid-search & $\textbf{21.4}_{.1}$ & $0.38_{.10}$ & 230 & 8k & 1 & 23.4 & 1\\
        & & TPE & $22.6_{.2}$ & $0.39_{.13}$ & 217 & 7.5k & 0.88 & 23.5 & 17\\
        & & Fixed values &  $24.9_{.2}$ & $0.38_{.08}$ & 99 & 5.1k & 0.46 & 16.1 & 0.49\\
        
        \cmidrule{2-10}
        
        & \multirow{5}{*}{{\rotatebox[origin=c]{90}{(MFCC)}}}
        & $\mathcal{W}_G$  & $\textbf{68.4}_{.1}$ & $5.1_{.08}$ & \textbf{99} & 8k & \textbf{0.76} & \textbf{8.2} & 1.13\\
        & & $\mathcal{W}_{HM}$ + $\mathcal{D}_s$ &  $70.5_{.2}$ & $6.3_{.14}$ & 114 & \textbf{7.1k} & \textbf{0.76} & 8.8 & 1.19\\
        & & Grid-search & $70.0_{.1}$ & $\textbf{4.4}_{.07}$ & 220 & 8k & 1 & 12.7 & 1 \\
        & & TPE & $70.8_{.2}$ & $4.9_{.18}$ & 247 & 8k & 1.0 & 12.7 & 14\\
        & & Fixed values & $70.7_{.1}$ & $8.9_{0.19}$ & 99 & 8k & 0.76 & 8.2 & 1.13  \\
        
        \midrule
        \multicolumn{3}{c|}{\textbf{TBI detection}} & \multicolumn{2}{c|}{\textbf{BA} $(\%_{\pm std})$} & \\
        \midrule

        \multirow{5}{*}{{\rotatebox[origin=c]{90}{\textbf{cGRU}}}} & \multirow{5}{*}{{\rotatebox[origin=c]{90}{(Raw audio)}}}
        & $\mathcal{W}_{HM}$ &  \multicolumn{2}{c|}{$86.53_{1.3}$} & \textbf{2.89s} & 8k & 0.79 & \textbf{3.7}s &1.05\\
        & & $\mathcal{W}_{HM}$ + $\mathcal{D}_s$ &  \multicolumn{2}{c|}{$\textbf{87.12}_{1.4}$} & 3.14s & \textbf{6.2k }& \textbf{0.74} & \textbf{3.7}s &1.02\\
        
        & & Grid-search &  \multicolumn{2}{c|}{$83.82_{1.4}$} & 4s & 8k & 1 & 4.6s & 1 \\
        & & TPE &  \multicolumn{2}{c|}{$81.90_{1.9}$} & 3.94s & 8k & 1 & 4.6s & 18\\
        & & Fixed values &  \multicolumn{2}{c|}{$82.62_{1.1}$} & 3.14s & 6.2k & 0.74 & 3.7s & 0.78\\
        
        \bottomrule
         \multicolumn{10}{p{\linewidth}}{MAC and training time are reported as a ratio to Grid-search.}
    \end{tabular}}
    \label{table:speaker_recognition}

\end{table}

\subsection{TBI detection}
As reported in Table~\ref{table:speaker_recognition}, the best TBI detection BA is obtained using 3.14s of speech sampled at 12.4 kHz, improving the grid-search result by 3.9\%. The training time used to tune the model is significantly lower than TPE and is competitive with grid-search that trained DNN in parallel. Energy consumption at inference is expected to reduce by 26\% compared to baseline. Similar to speaker recognition results, windowing and $\mathcal{D}_s$ layers allow the DNN to learn from different lengths and sampling rates of speech, which provides a better detection BA.

\section{Conclusion}
\label{sec:conclusion} 
DNN-based speech processing has the potential for impact but currently has high energy consumption, limiting the mobile deployment of state-of-the-art methods. This study proposed learning an optimal speech length and sampling rate using a masking function during DNN back-propagation, which reduces the energy consumption of speech acquisition and DNN inference. 
Our evaluation demonstrates that learning speech format using an end-to-end model outperforms tuning window length and sampling rate as hyperparameters. As estimated using the MAC metric, the power consumption used for inference is reduced by up to 73\% and 26\% for the speaker recognition and TBI detection tasks, respectively, while maintaining high accuracy. Beyond the speaker recognition and TBI detection tasks, the proposed method is also broadly applicable to other speech tasks.

Our proposed method has the limitation of requiring subsequent DNN layers to operate on a tensor with a dynamic temporal dimension. The fully-connected DNN layer was modified to have a flexible input size, which may create inconsistent losses across training epochs. In the future, we plan to investigate additional speech processing parameters, such as hop size, to further reduce energy consumption of mobile speech processing.

\flushend

\bibliographystyle{IEEEtran}
\bibliography{reference}

\end{document}